# A randomized, efficient algorithm for 3SAT

Cristian Dumitrescu
Independent mathematician, Kitchener, Canada.
cristiand43@gmail.com

**Abstract.** In this paper, we study an extension of Schöning's algorithm [Schöning, 1991] for 3SAT, based on the concept of truth valuation, which is a generalization of the concept of truth assignment. We also formulate a conjecture and present strong arguments that a version of the algorithm is polynomial.

## 1 Introduction

The importance of efficiently solving NP – complete problems is due to the fact that this would imply that all the other problems belonging to the NP class can be efficiently solved in a constructive manner (the algorithms can be generated for all of them through polynomial - time reductions).

It is well known that 3SAT is NP-complete.

**Proposition 1.** (see [Papadimitriou, 94] and [Hopcroft, 1979]). 3SAT is NP-complete.

**Proof.** For the proof, see (see [Papadimitriou, 94] and [Hopcroft, 1979]).

We will first present a well known randomized algorithm for 3SAT. This is Schöning's algorithm from 1991 (see [Schöning, 1991]). We also note that Papadimitriou also discussed a similar algorithm for 2SAT in 1991 (see [Papadimitriou, 1991]).

It is known that Papadimitriou's algorithm finds a solution in quadratic time with high probability for 2SAT. Only exponential bounds are known for Schöning's algorithm for 3SAT.

Stochastic local search algorithms played an important role, related to SAT solvers. This includes the random walk algorithm of Papadimitriou for 2SAT, and Schöning's extension of this algorithm for 3SAT.

*Schöning's algorithm for 3SAT*

*Input: a 3SAT expression in $n$ variables.*

*Guess an initial truth assignment, uniformly at random.*

*Repeat $3 \cdot n$ times:*

> *If the expression is satisfied by the actual assignment, stop and accept.*
>
> *Let C be some clause not being satisfied by the actual assignment. Pick one of the three literals in the clause at random, and flip its truth value.*
>
> *Update.*

*Stop and reject, the expression is not satisfiable.*

Note that these algorithms involve random walks on the boundary of the truth assignment hypercube (the unit hypercube where each vertex represents a truth assignment). The fundamental and new idea in this article is to consider stochastic local search algorithms through the bulk (the interior) of the truth assignment hypercube (which becomes the truth valuation hypercube). This will become clear in section 2, when the concept of truth valuation will be introduced. We will formulkate a conjecture and present strong arguments that the algorithm presented in section 2 is polynomial (even for the worst case instances of 3SAT).

A different type of algorithm (this is a hill climbing algorithm, not directly related to Papadimitriou or Schöning's algorithms), but also based on the concept of truth valuations, is given in [A2].

Still another type of algorithm (the clustered Sparrow algorithm) is presented in [A4].

## 2 The extension of Schöning's algorithm

We assume that the reader is familiar with the random walk model (based on the Hamming distance to a solution) associated with the algorithm. This is discussed in detail in [Schöning, 1991], and [Papadimitriou, 1991].

**Definition 1.** In the following, M will be a natural number, in general smaller than n. A truth valuation $v(x)$ is a function from the set of variables $\{x_1, x_2, x_3, \dots\dots.x_n\}$ to the set of values $\left\{0, \frac{1}{M}, \frac{2}{M}, \frac{3}{M}, \dots\dots, \frac{M-1}{M}, 1\right\}$. Each variable will be assigned a rational number from the set $\{0, \frac{1}{M}, \frac{2}{M}, \frac{3}{M}, \ldots\ldots, \frac{M-1}{M}, 1\}$. We note the difference between a truth valuation and a truth assignment, where the truth assignment for a variable can be only 0 or 1. We also note that in general (as used for the truth valuation of a clause), a truth valuation is a real value in the interval $[0, 1]$.

**Definition 2.** Given two truth valuations for n variables $(v_1(x_1), v_1(x_2), v_1(x_3), \ldots\ldots v_1(x_n))$, and $(v_2(x_1), v_2(x_2), v_2(x_3), \ldots\ldots v_n(x_n))$, the Hamming distance between these truth valuations is by definition $d(v_1, v_2) = |v_1(x_1) - v_2(x_1)| + |v_1(x_2) - v_2(x_2)| + \cdots \ldots + |v_1(x_n) - v_2(x_n)|$.
The normalized Hamming distance is just the Hamming distance divided by the length of one step, which is $\frac{1}{M}$.

**Definition 3.** The truth valuation of $x \vee y$ is $v(x \vee y) = v(x) + v(y) - v(x) \cdot v(y)$. The valuation of the negation of a variable $\neg x$ is $v(\neg x) = 1 - v(x)$. It is clear that if we are given a truth valuation for the n variables that appear in a 3SAT expression, then for any 3SAT clause we can calculate the truth valuation of that clause $v(x \vee y \vee z) = v(x) + v(y) + v(z) - v(x) \cdot v(y) - v(x) \cdot v(z) - v(y) \cdot v(z) + v(x) \cdot v(y) \cdot v(z)$. The truth valuation of the whole 3SAT expression is the product of the truth valuations of all the clauses.

*Extension of Schöning's algorithm for 3SAT*

*Input: a 3SAT expression in $n$ variables.*
*Initialize all variables with the truth valuation $\frac{1}{2}$ (or following a uniform distribution on the states of the class $S_0$, described below).*

*Repeat $4 \cdot n^2 \cdot M^2$ times:*

> *If the current truth valuation is a satisfying truth assignment, then stop and return the satisfying truth assignment.*
>
> *Let C be some clause that has a minimal truth valuation. If there are more than one, then pick one at random. . Choose one of the literals in the clause at random ,and if its current truth valuation is not 0 or 1, then with equal probability $\frac{1}{2}$, increase or decrease the truth valuation of the corresponding variable by $\frac{1}{M}$ . If the current truth valuation of the variable is 0, then with probability 1 increase its truth valuation with $\frac{1}{M}$. If the current truth valuation of the variable is 1, then with probability 1 decrease its truth valuation with $\frac{1}{M}$.*
>
> *Update.*

*Stop and reject, the expression is not satisfiable.*

The purpose of this paper is to prove that this algorithm finds a solution (if it exists) with high probability. We will need some preparatory propositions.

**Proposition 2**. We consider the case of one dimensional symmetric random walks with one unit steps (like the coin tossing game). For fixed t, the probability that the first passage through r occurs before epoch $t \cdot r^2$ tends to $\sqrt{\frac{2}{\pi}} \cdot \int_{\frac{1}{\sqrt{t}}}^{\infty} e^{-\frac{1}{2}\cdot s^2} ds = 2\,(1 - R(\frac{1}{\sqrt{t}}))$ , as $r \to \infty$ , where R is the normal distribution.

**Proof.** For the proof, see [Feller, 1968], chapter 3.

This means that the waiting time for the first passage through r increases with the square of r. The expected traveled distance in $r^2$ steps is r.

We note that similar to the random walk model ([Schöning, 1991], and [Papadimitriou, 1991]), we have here a Markov chain. At any moment, if the current truth valuation of the variable is not 0 or 1, then the Hamming distance from the current truth valuation (for all the n variables) to the truth assignment solution (if it exists) can increase or decrease by $\frac{1}{M}$ with probability $\frac{1}{2}$. But at certain moments, the Hamming distance will increase or decrease with probability 1 (when the current truth valuation gives the corresponding variable the value 0 or 1). We do not have a pure random walk, as in the original models. We will call this the primary Markov chain, associated to the Hamming distance to a solution. The states of the primary Markov chain indicate the Hamming distance towards a solution.

We also note that we have n one dimensional symmetric random walks with reflecting barriers at 0 and 1 (each such random walk is associated to the truth valuation of a single variable). We will call these the secondary random walks with reflecting barriers. The states of the n secondary random walks with reflecting barriers indicate the current truth valuation of the corresponding variable.

**Proposition 3.** The secondary random walks with reflecting barriers have stationary distributions.

**Proof.** In the secondary random walks with reflecting barriers, when we say that we are in state $\frac{i}{M}$, we mean that the current truth valuation of the corresponding variable is $\frac{i}{M}$. The probability transition matrix A for each secondary random walk is defined by the transition probabilities:

$p\left(\frac{i}{M}, \frac{j}{M}\right) = 0$ if $|i - j| > 1$ , for any values of $i, j \in \{0, 1, 2, 3, \dots\dots.., M\}$ .

$p\left(\frac{i}{M}, \frac{i}{M}\right) = 0$ , for any values of $i \in \{0, 1, 2, 3, \dots\dots.., M\}$ .

$p\left(\frac{i}{M}, \frac{i+1}{M}\right) = p\left(\frac{i}{M}, \frac{i-1}{M}\right) = \frac{1}{2}$ , for any values of $i \in \{1, 2, 3, \dots\dots.., M - 1\}$

$p\left(0, \frac{1}{M}\right) = 1$

$p\left(1, \frac{M-1}{M}\right) = 1$ .

The probability transition matrix is:

$$A = \begin{pmatrix} 0 & 1 & 0 & 0 & \cdots & 0 & 0 & 0 & 0 \\ \frac{1}{2} & 0 & \frac{1}{2} & 0 & \cdots & 0 & 0 & 0 & 0 \\ 0 & \frac{1}{2} & 0 & \frac{1}{2} & \cdots & 0 & 0 & 0 & 0 \\ 0 & 0 & \frac{1}{2} & 0 & \cdots & 0 & 0 & 0 & 0 \\ \vdots & \vdots & \vdots & \vdots & \ddots & \vdots & \vdots & \vdots & \vdots \\ 0 & 0 & 0 & 0 & \cdots & 0 & \frac{1}{2} & 0 & 0 \\ 0 & 0 & 0 & 0 & \cdots & \frac{1}{2} & 0 & \frac{1}{2} & 0 \\ 0 & 0 & 0 & 0 & \cdots & 0 & \frac{1}{2} & 0 & \frac{1}{2} \\ 0 & 0 & 0 & 0 & \cdots & 0 & 0 & 1 & 0 \end{pmatrix}$$

If we write ( $\pi(0), \pi\left(\frac{1}{M}\right), \pi\left(\frac{2}{M}\right), \ldots \ldots, \pi(1)$ ) for the stationary distribution, it will be the solution of the system of equations (this is a $1 \times (M+1)$ row matrix multiplied by a $(M+1) \times (M+1)$ square matrix):

$$\left( \pi(0) \;\; \pi\left(\frac{1}{M}\right) \;\; \pi\left(\frac{2}{M}\right) \;\; \ldots \ldots . \pi(1) \right) \cdot A = \left( \pi(0) \;\; \pi\left(\frac{1}{M}\right) \;\; \pi\left(\frac{2}{M}\right) \;\; \ldots \ldots . \pi(1) \right) .$$

When we solve this system of equations, we will find:

$$\pi(0) = \pi(1) = \frac{1}{2M}$$

$$\pi\left(\frac{1}{M}\right) = \pi\left(\frac{2}{M}\right) = \pi\left(\frac{3}{M}\right) = \cdots \ldots . = \pi\left(\frac{M-1}{M}\right) = \frac{1}{M} .$$

This stationary distribution is an equilibrium distribution. We need some results related to the convergence towards a stationary distribution for periodic chains.

**Definitions 4.** Two states in a Markov chain are in the same class if we can go from one to another in a finite number of steps with positive probability. A Markov chain is said to be irreducible if all states are in the same class. A recurrent state is one for which the probability that the Markov chain will return to it after a finite time is one. Assume that state $j$ is a recurrent state. Let $I = \{n \geq 1;\ p_n(j,j) > 0\}$ , and let $d$ be the greatest common divisor of $I$. We call $d$ the period of state $j$ . A chain with period 1 is said to be aperiodic. All states in the same recurrent class have the same period.

**Proposition 4.** Suppose the Markov chain M is irreducible, recurrent and all states have period d. Fix x in S (the set of states), and for each y in S, we let $K_y = \{n \geq 1 ;\ p_n(x,y) > 0\}$. Then the following statements are true:

(i) There is an $r_y \in \{0, 1, \ldots \ldots, d-1\}$ so that if $n \in K_y$, then $n = r_y \bmod d$.

(ii) Set $S_r = \{y ;\ r_y = r\}$ for $0 \leq r < d$. If $y \in S_i$ and $z \in S_j$ and $p_n(y,z) > 0$, then we have $n = (j-i) \bmod d$.

(iii) $S_0, S_1, \ldots \ldots \ldots, S_{d-1}$ are irreducible classes for $M^d$ (the d – step chain constructed from M), and all states have period 1 (this is called a cyclic decomposition of the state space).

**Proof.** For the proof, see [Durrett, 2010], chapter 6.

**Theorem 1.** (convergence theorem for periodic chains). Suppose the Markov chain M is irreducible, has a stationary distribution $\pi$, and all states have period d. Let $x \in S$, and let $S_0, S_1, \ldots \ldots \ldots, S_{d-1}$ be a cyclic decomposition of the state space with $x \in S_0$. If $y \in S_r$, then $\lim_{m \to \infty} p_{md+r}(x,y) = d \cdot \pi(y)$.

**Proof.** For the proof, see [Durrett, 2010], chapter 6.

**Examples.** We will look at the secondary random walks with reflecting barriers for different values of M.

We note that the case $M = 1$ corresponds to the original version of Schöning's algorithm. In this case the transition probability matrix is $A = \begin{pmatrix} 0 & 1 \\ 1 & 0 \end{pmatrix}$. We see that $A^2 = I$ (the identity matrix). We have two states, state 0 (when the truth valuation of the variable is 0), and state 1 (when the truth valuation is 1). We have the transition probabilities $p(0,1) = p(1,\ 0) = 1$. The stable distribution of this chain is $\pi(0) = \pi(1) = \frac{1}{2}$. The chain is periodic with period $d = 2$. The cyclic decomposition of the state space is $S_0 = \{0\},\ S_1 = \{\,1\,\}$. In this case, we have $p_{2m}(0,0) = p_{2m+1}(0,1) = 1$. If we start in state 0, an example of a path taken by this chain is $1, 0, 1, 0, 1, 0, \ldots \ldots \ldots$. We see that at any odd time, the chain is in state 1, and at every even time the chain is in state 0. If we choose a state at random from a given path, we recover the stationary distribution $\pi(0) = \pi(1) = \frac{1}{2}$.

We will now consider the case $M = 2$ in greater detail. In this case the transition probability matrix is $A = \begin{pmatrix} 0 & 1 & 0 \\ \frac{1}{2} & 0 & \frac{1}{2} \\ 0 & 1 & 0 \end{pmatrix}$. We see that $A^2 = \begin{pmatrix} \frac{1}{2} & 0 & \frac{1}{2} \\ 0 & 1 & 0 \\ \frac{1}{2} & 0 & \frac{1}{2} \end{pmatrix}$, and we also see that $A^3 = A$. We have three states, state 0 (when the truth valuation of the variable is 0), state $\frac{1}{2}$ (when the truth valuation is $\frac{1}{2}$), and state 1 (when the truth valuation is 1). We have the transition probabilities $p\left(0,\frac{1}{2}\right) = p\left(1,\frac{1}{2}\right) = 1$, and $p\left(\frac{1}{2},0\right) = p\left(\frac{1}{2},1\right) = \frac{1}{2}$. The stable distribution of this chain is $\pi(0) = \pi(1) = \frac{1}{4}$, $\pi\left(\frac{1}{2}\right) = \frac{1}{2}$. The chain is periodic with period $d = 2$. The cyclic decomposition of the state space is $S_0 = \{0,1\}$, $S_1 = \{\frac{1}{2}\}$. The convergence theorem for periodic chains tells us that $\lim_{m\to\infty} p_{2m}(0,0) = \lim_{m\to\infty} p_{2m}(0,1) = \frac{1}{2}$, and $\lim_{m\to\infty} p_{2m+1}\left(0,\frac{1}{2}\right) = 1$. If we start in state 0, an example of a path taken by this chain is $\frac{1}{2}, 0, \frac{1}{2}, 0, \frac{1}{2}, 1, \frac{1}{2}, 0, \frac{1}{2}, 1, \frac{1}{2}, 1, \frac{1}{2}, \ldots\ldots\ldots\ldots\ldots\ldots$ We see that at any odd time, the chain is in state $\frac{1}{2}$, and at every even time the chain is in state 0 or 1 with equal probability. If we choose a state at random from a given path, we recover the stationary distribution $\pi(0) = \pi(1) = \frac{1}{4}$, $\pi\left(\frac{1}{2}\right) = \frac{1}{2}$.

The case $M = 4$ is considered in [A3].

**Proposition 5.** In the general case, the period is $d = 2$. If $M$ is an even number, then the cyclic decomposition of the state space is

$S_0 = \left\{0, \frac{2}{M}, \frac{4}{M}, \frac{6}{M}, \ldots\ldots.., 1\right\}$, $S_1 = \left\{\frac{1}{M}, \frac{3}{M}, \frac{5}{M}, \ldots\ldots.., \frac{M-1}{M}\right\}$. We also have

$$\lim_{m\to\infty} p_{2m}(0,0) = \lim_{m\to\infty} p_{2m}(0,1) = \frac{1}{M} \qquad (1)$$

$$\lim_{m\to\infty} p_{2m}\left(0,\frac{2k}{M}\right) = \frac{2}{M}, \text{ for } 1 \le k \le \frac{M-2}{2} \qquad (2)$$

$$\lim_{m\to\infty} p_{2m+1}\left(0,\frac{2k+1}{M}\right) = \frac{2}{M}, \text{ for } 0 \le k \le \frac{M-2}{2} \qquad (3)$$

If $M$ is an odd number, then the cyclic decomposition of the state space is

$S_0 = \left\{0, \frac{2}{M}, \frac{4}{M}, \frac{6}{M}, \ldots\ldots.., \frac{M-1}{M}\right\}$, $S_1 = \left\{\frac{1}{M}, \frac{3}{M}, \frac{5}{M}, \ldots\ldots.., 1\right\}$. We also have

$$\lim_{m\to\infty} p_{2m}(0,0) = \lim_{m\to\infty} p_{2m+1}(0,1) = \frac{1}{M} \qquad (4)$$

$$\lim_{m\to\infty} p_{2m}\left(0,\frac{2k}{M}\right) = \frac{2}{M}, \text{ for } 1 \le k \le \frac{M-1}{2} \qquad (5)$$

$$\lim_{m\to\infty} p_{2m+1}\left(0,\frac{2k+1}{M}\right) = \frac{2}{M}, \text{ for } 0 \le k \le \frac{M-3}{2} \qquad (6)$$

**Proof.** The proof is immediate from proposition 4 and theorem 1.

**Definition 5.** If the current truth valuation of a variable is 0, but the satisfying truth assignment for this variable is 1, or if the current truth valuation of a variable is 1, but the satisfying truth assignment for this variable is 0, then we say that this will be a positive reflection. In this case, in the primary Markov chain, the Hamming distance to a solution will decrease with probability 1.

If the current truth valuation of a variable is 0, and the satisfying truth assignment for this variable is also 0, or if the current truth valuation of a variable is 1, and the satisfying truth assignment for this variable is also 1, then we say that this will be a negative reflection. In this case, in the primary Markov chain, the Hamming distance to a solution will increase with probability 1.

The states $\frac{i}{M}$ , with $1 \leq i \leq M-1$ will be called intermediate states.

In the algorithm, choosing a variable to change its truth valuation is equivalent to running one of the secondary random walks (the one associated to the variable chosen) for one step, and that is equivalent to one step performed by the primary chain.

**Conjecture**. At least for large values of M (comparable to n), the algorithm above will find a solution to a 3SAT problem (if it exists) in polynomial time, with high probability.

**Arguments in favor of the conjecture.** We have to study the ergodic properties of the primary Markov chain described above. For this purpose, we use proposition 5, relations (1) – (6). We have to consider the two step Markov chain, on one of the cyclic classes $S_0$ , or $S_1$ . As an example, for the case $M = 2$ , we have $S_0 = \{0, \ 1\}$ , $S_1 = \left\{\frac{1}{2}\right\}$ . For the two step chain, we have the probabilities : $p\left(\frac{1}{2} \to 0 \to \frac{1}{2}\right) = p\left(\frac{1}{2} \to 1 \to \frac{1}{2}\right) = \frac{1}{2}$ . We can couple our chain with a symmetric random walk. For each variable, starting in the state $\frac{1}{2}$, the chain visits the correct truth assignment, or the incorrect one with the same probability $\frac{1}{2}$. We know from proposition 2 that for a symmetric random walk, the expected traveled distance after $N^2$ steps is of order of magnitude $N$ . In the context of our two step chain, this means that after $N^2$ steps, there might be a sufficiently large excess of variables that visit the correct truth assignment. Of course, this is just a heuristic argument, but the study of the ergodic properties of the primary Markov chain could lead to the conclusion that its behavior resembles the properties of a symmetric random walk, and in this case we know that a quadratic number of steps is sufficient , in order to find a solution with high probability.

**Observation 1.** There is an interesting geometrical interpretation. The $n$ random walks with reflecting barriers (one associated to each variable) generate a random walk inside the truth valuation hypercube. The truth valuation hypercube has a side length of 1 (but we keep in mind that M is the minimum number of steps for a variable to change its truth assignment from 0 to 1 or from 1 to 0), and its main diagonal has the length $\sqrt{n}$ . We could assume that the random walk starts at $(0, 0, 0, \dots \dots, 0)$ , and it reaches the vertex of the hypercube that represents the truth assignment solution . As a result of the construction, this $n-$ dimensional random walk never exits the interior of the hypercube, even if at times it can take place on its boundary. Also, this is not a perfectly symmetric random walk, because at each step, it follows the direction of the variables involved in minimal clauses (clauses with a minimal truth valuation). This will give our random walk a general "drift" direction towards the solution vertex of the truth valuation hypercube, and we also know that the distance that has to be travelled is less than $\sqrt{n}$. Since the distance travelled scales up as the square root of the number of steps performed by the algorithm, this means that a quadratic number of steps (up to a factor that depends on $M$ ) are far more than sufficient, in order for the algorithm to hit a solution. I emphasize that this is only a heuristic proof, and more work needs to be done in this direction.

**Observation 2.** We also note that this is only one version of this type of algorithm. Other versions or improvements are possible. The original Schöning algorithm ( $M = 1$ ), relies only on reflections. In the version of the algorithm presented here, we try to minimize the effect of reflections, and rely instead on intermediate steps to drive the primary Markov chain towards a solution.

## 3 Discussion and conclusions

For general implications, related to efficiently solving NP – complete problems, see [Fortnow, 2013]. An interesting application is related to the problem of automated theorem proving using an efficient algorithm for NP – complete problems (see [A1]). The impact of this type of algorithm in mathematics is obvious. All unsolved problems at the present time (if they allow proofs that are not too long, within the axiomatic system that we work with, for example, with the information content equivalent to less than 10000 journal pages) will be quickly solved. If NP complete problems can be solved efficiently, better AI system can be designed. In industry, transportation problems, logistics, and manufacturing planning will be significantly improved. In medicine, these algorithms will lead to the design of better drugs and treatments (protein folding and its relation to NP complete problems). Also the crypto community must redesign their encoding and decoding procedures (and public key cryptography must be replaced). Basically, any field of activity that requires intellectual effort can benefit from these types of algoritms. The main purpose of this paper is to open new ways to approach this interesting problem.

**Aknowledgements.**

I emphasize that Professor Fortnow's book ([Fortnow, 2013]) was a great source of inspiration and motivation that led me to seriously consider this problem. Also, I received some feedback related to previous versions of my work from Professor Uwe Schöning , Professor Lance Fortnow, and Professor C. Papadimitriou. I also discussed some of the arguments presented above with Eric Demer , graduate student at University of California, Santa Barbara, who also corrected a previous error that I had in my work. The clustered Sparrow algorithm from [ A4] was developed in collaboration with Dr. Anastasia-Maria Leventi-Peetz (BSI, Germany). I am grateful to all. In the same time, I take full responsibility for any omissions present in my work.

## Appendices

**A1.** Godel's letter to John von Neumann. In his letter, Godel writes:
"One can obviously easily construct a Turing machine, which for every formula F in first order predicate logic and every natural number n, allows one to decide if there is a proof of F of length n (length = number of symbols). Let $\psi(F,n)$ be the number of steps the machine requires for this and let $\varphi(n) = max_F \psi(F,n)$. The question is how fast $\varphi(n)$ grows for any optimal machine" (see [Godel, 1956].

Now we consider this. In [Hopcroft, 1979], we have theorem 13.1, at page 325 (a version of the Cook - Levin theorem), where it is proved that for each Turing machine (deterministic or nondeterministic) M that is time bounded by a polynomial $p(n)$, a log-space algorithm exists, that takes as input a string x and produces a Boolean expression $E_x$ that is satisfiable if and only if M accepts x.

This means that the process of seeking a proof (of reasonable length) of a mathematical statement can be completely automatized. With the algorithm presented in this paper, Godel's vision can be made reality.

**A2.** We are given a 3SAT expression E. A positive literal represents the variable itself, a negative literal represents the negation of a variable. For each occurrence of a variable x (in a positive or negative literal) within the expression we introduce a new variable $x_i$ , and form the modified expression E*. We also have to include a conjunction of clauses that state that the variables $x_i$ have the same truth assignment. The conjunction $(x_i \vee \urcorner x_j) \wedge (\urcorner x_i \vee x_j)$ has the truth value 1 if and only if the variables $x_i$ and $x_j$ have the same truth value. In general, the variables $x_1, x_2, x_3, \ldots\ldots.., x_m$ have the same truth value if and only if the conjunction $(x_1 \vee \urcorner x_2) \wedge (x_2 \vee \urcorner x_3) \wedge (x_3 \vee \urcorner x_4) \wedge \ldots\ldots\ldots. \wedge (x_{m-1} \vee \urcorner x_m) \wedge (x_m \vee \urcorner x_1)$ has the truth value 1. In the modified expression E*, each variable appears in at most three clauses, either in two positive literals and one negative, or in two negative literals and one positive.

In other words, each variable x appears only in a three variable clause $( x \vee \alpha \vee \beta )$ (or it could be $( \urcorner x \vee \alpha \vee \beta )$ ) , and in a couple of two variable clauses $( x \vee \urcorner \gamma)$ and $( \urcorner x \vee \theta)$. Each variable appears only in the conjunction of these clauses $( x \vee \alpha \vee \beta ) \wedge (x \vee \urcorner \gamma) \wedge (\urcorner x \vee \theta)$ . We now consider the valuation of these clauses , in terms of the valuations of each variable that appears in these clauses (we keep in mind that the truth valuation of a variable is a real number in the interval $[0, 1]$, in this appendix). We calculate the valuation of each of the three clauses where the variable x appears.

$$v( x \vee \alpha \vee \beta) = v(x) + v(\alpha) + v(\beta) - v(x) \cdot v(\alpha) - v(x) \cdot v(\beta) - v(\alpha) \cdot v(\beta) + v(x) \cdot v(\alpha) \cdot v(\beta) .$$

$$v(x \vee \urcorner \gamma) = v(x) + v(\urcorner \gamma) - v(x) \cdot v(\urcorner \gamma)$$

$v(\urcorner x \vee \theta) = v(\urcorner x) + v(\theta) - v(\urcorner x) \cdot v(\theta)$

We also keep in mind that $v(\urcorner x) = 1 - v(x)$ and $v(\urcorner \gamma) = 1 - v(\gamma)$.

The function $f(x) = v(x \vee \alpha \vee \beta) \cdot v(x \vee \urcorner \gamma) \cdot v(\urcorner x \vee \theta)$ becomes a cubic polynomial in the truth valuation of the variable x, if we look at the valuations of $\alpha$, $\beta$, $\gamma$, and $\theta$ as fixed parameters. We can then calculate the valuation of the variable x, such that the expression f(x) takes a maximum value in the interval $[0, 1]$, and change the valuation of x to that value. We remember that we want to maximize the truth valuations of all the clauses in the expression E*, until they all become 1. We can repeat this process for all the variables involved in the expression E*, over and over again, until all clauses (and thus the whole expression E*) has maximum truth valuation 1. The problem is that we can get stuck in a local maximum, even if the truth valuation of every clause is not the maximum 1. In that case, we can reinitialize the truth valuations of the variables at random and start all over again, until we enter the basin of attraction of the global maximum, that we are seeking. At each step, the truth valuation of the whole expression E* can only increase or stay the same.

**A3.** If in a solution truth assignment, a variable is given the truth assignment 0, then the truth assignment 1 is a positive reflection point, and the truth assignment 0 is a negative reflection point, for the random walk with reflecting barriers associated to this variable. If in a solution truth assignment, a variable is given the truth assignment 1, then the truth assignment 0 is a positive reflection point, and the truth assignment 1 is a negative reflection point, for the random walk with reflecting barriers associated to this variable.

We consider the random walks with reflecting barriers associated to each variable. We consider the case $M = 4$. In this case, we have the states $0, \frac{1}{4}, \frac{1}{2}, \frac{3}{4}, 1$, and the probability transition matrix is:

$$A = \begin{pmatrix} 0 & 1 & 0 & 0 & 0 \\ \frac{1}{2} & 0 & \frac{1}{2} & 0 & 0 \\ 0 & \frac{1}{2} & 0 & \frac{1}{2} & 0 \\ 0 & 0 & \frac{1}{2} & 0 & \frac{1}{2} \\ 0 & 0 & 0 & 1 & 0 \end{pmatrix}$$

It can be easily proved that we have the following relations, for $k \geq 1$.

$$A^{2k-1} = \begin{pmatrix} 0 & \frac{2^{k-1}+1}{2^k} & 0 & \frac{2^{k-1}-1}{2^k} & 0 \\ \frac{2^{k-1}+1}{2^{k+1}} & 0 & \frac{1}{2} & 0 & \frac{2^{k-1}-1}{2^{k+1}} \\ 0 & \frac{1}{2} & 0 & \frac{1}{2} & 0 \\ \frac{2^{k-1}-1}{2^{k+1}} & 0 & \frac{1}{2} & 0 & \frac{2^{k-1}+1}{2^{k+1}} \\ 0 & \frac{2^{k-1}-1}{2^k} & 0 & \frac{2^{k-1}+1}{2^k} & 0 \end{pmatrix}$$

$$A^{2k} = \begin{pmatrix} \frac{2^{k-1}+1}{2^{k+1}} & 0 & \frac{1}{2} & 0 & \frac{2^{k-1}-1}{2^{k+1}} \\ 0 & \frac{2^k+1}{2^{k+1}} & 0 & \frac{2^k-1}{2^{k+1}} & 0 \\ \frac{1}{4} & 0 & \frac{1}{2} & 0 & \frac{1}{4} \\ 0 & \frac{2^k-1}{2^{k+1}} & 0 & \frac{2^k+1}{2^{k+1}} & 0 \\ \frac{2^{k-1}-1}{2^{k+1}} & 0 & \frac{1}{2} & 0 & \frac{2^{k-1}+1}{2^{k+1}} \end{pmatrix}$$

These probability transitions govern all the odd and even steps of the random walk with reflecting barriers associated to the variables. We see that if a variable starts in a positive reflection point, then at even steps, it is slightly more likely that the associated random walk with reflecting barriers will hit the positive reflection point, rather than the negative reflection point. It is also more likely that at odd steps, the random walk will hit points closer to the positive reflection point, rather than closer to the negative reflection point. Without restricting generality, if we assume that most of the variables start in positive reflection points (for example, we can initialize all the variables with 0 or we can initialize all the variables with the truth value 1), it is then likely that overall, the positive reflection points will be hit more often than the negative reflection points, but this also depends on the structure of the 3SAT expression, and the complex dynamics of the primary Markov chain. It is worth noticing that if all the variables are initialized with the truth valuation $\frac{1}{2}$ , then at even steps, it is as likely that the random walk with reflecting barriers (associated to any variable) will hit a positive reflection point, as it is likely that it will hit a negative reflection point (in both cases the probability is $\frac{1}{4}$ ).

**A4. The clustered Sparrow algorithm**

We start from a 3SAT expression E, and we perform the conversion to the clustered SAT expression E*, where each variable appears in at most 3 clauses. For a clear presentation of the procedure, see proposition 9.3, page 183, in [Papadimitriou, 1994].

We note that if the 3SAT expression E has n variables and c clauses, then the clustered expression E* will contain $3c$ variables, and $4c$ clauses, out of which $c$ will be three variable clauses and $3c$ will be two variable clauses.

A cluster associated to a variable x, can be of two types.
Type 1 cluster:
(x ∨ y ∨ z)∧(x ∨ ˥u)∧(˥x ∨ w)

Here y and z are two literals (they can be positive or negative), and we write $˥x$ for the negative literal that represents the negation of x.

Type 2 cluster:
(˥x ∨ y ∨ z)∧(x ∨ ˥u)∧(˥x ∨ w)

In the clusters above, we will call x the primary variable associated to the cluster, and the other variables are called secondary (but they are primary variables as seen in their own clusters). In a cluster, the three variable clause, and the two variable clause where the primary variable is under the same literal (both positive or bot negative) are called primary clauses. The two variable clause in the cluster in which the primary variable is under an opposite literal (compared to the three variable clause), is called the secondary clause. So a cluster has two primary clauses and a secondary clause.

The most general form of a cluster can be written in the form:
(x ∨ a)∧(x ∨ b)∧(˥x ∨ c)
In this case, b, and c can be positive or negative literals, and a is a disjunction of literals, $a =$ y ∨ z , b = ˥u , c = w , and x can be a positive or negative literal
We construct the truth table of the cluster.

| | x | ˥x | a | b | c | $x \vee a$ | $x \vee b$ | $˥x \vee c$ | $0 \to 1$ | $1 \to 0$ |
|---|---|---|---|---|---|---|---|---|---|---|
| 1 | 0 | 1 | 0 | 0 | 0 | 0 | 0 | 1 | +1 | |
| 2 | 1 | 0 | 0 | 0 | 0 | 1 | 1 | 0 | | -1 |
| 3 | 0 | 1 | 1 | 0 | 0 | 1 | 0 | 1 | 0 | |
| 4 | 1 | 0 | 1 | 0 | 0 | 1 | 1 | 0 | | 0 |
| 5 | 0 | 1 | 0 | 1 | 0 | 0 | 1 | 1 | 0 | |
| 6 | 1 | 0 | 0 | 1 | 0 | 1 | 1 | 0 | | 0 |
| 7 | 0 | 1 | 0 | 0 | 1 | 0 | 0 | 1 | +2 | |
| 8* | 1 | 0 | 0 | 0 | 1 | 1 | 1 | 1 | | |
| 9* | 0 | 1 | 1 | 1 | 0 | 1 | 1 | 1 | | |
| 10 | 1 | 0 | 1 | 1 | 0 | 1 | 1 | 0 | | +1 |
| 11 | 0 | 1 | 1 | 0 | 1 | 1 | 0 | 1 | +1 | |
| 12* | 1 | 0 | 1 | 0 | 1 | 1 | 1 | 1 | | |
| 13 | 0 | 1 | 0 | 1 | 1 | 0 | 1 | 1 | +1 | |
| 14* | 1 | 0 | 0 | 1 | 1 | 1 | 1 | 1 | | |
| 15* | 0 | 1 | 1 | 1 | 1 | 1 | 1 | 1 | | |
| 16* | 1 | 0 | 1 | 1 | 1 | 1 | 1 | 1 | | |

We note that the assignments marked with * satisfy the cluster (all clauses in the cluster are satisfied). The flips $0 \rightarrow 1$ and $1 \rightarrow 0$ refer to the primary variable x, when all the other literals do not change (under the current truth assignment). The last two columns represent the make minus break quantity $\Delta = \text{make} - \text{break}$ , where "make" represents the number of clauses that change from unsatisfied to satisfied, as a result of the flip, and "break" represents the number of clauses that change from satisfied to unsatisfied, as a result of the flip. The algorithm only touches unsatisfied clauses (or clusters), so the satisfying assignments do not have a Δ value in the corresponding column (a possible exception is between assignments 15* and 16*, since the cluster is left satisfied and $\Delta = 0$ ).

We will call a positive flip, a flip that has the value of delta strictly positive $\Delta > 0$.
We will call a negative flip, a flip that has the value of delta strictly negative $\Delta < 0$ .
We will call a null flip, a flip that has the value of delta 0, $\Delta = 0$ .

We note that all flips are either positive or null, with one exception, the second assignment in the table (the delta value marked in red). Also, we note that if at a given moment, a negative flip occurs, we see that the cluster is still unsatisfied, allowing for a positive flip later. Thus, at later stages, the flip will be reversed with high probability, if the change of the secondary variables in the cluster do not make the cluster satisfied before that.

**Lemma.** If the clustered expression E* contains clusters with any of the primary clauses unsatisfied, then there are at least some null flips available (if not positive flips).

**Proof.** If a primary clause is unsatisfied, then the literal (positive or negative) associated to the primary variable of the cluster has truth value 0. When we flip the primary variable, and make this unsatisfied primary clause satisfied, the other primary clause cannot become unsatisfied as a result of this flip. Only the secondary clause in this cluster could become unsatisfied as a result of this flip (but not necessarily). As a consequence, the make – break value of this flip is greater or equal to zero $\Delta \geq 0$ .

We have to construct a probability function that will determine the type of flip to be chosen, at each step. Let's assume that at a given moment, we have available $N_1$ negative flips, $N_2$ null flips, and $N_3$ positive flips (where $N_1 + N_2 + N_3$ represents the number of variables that have unsatisfied clusters). We want to assign a higher probability to the positive and null flips, and a lower probability to the negative flips. We choose a small positive number α. If we assign the probability $\frac{\alpha}{N_1}$, to choose a particular negative flip, then the probability to choose any negative flip is α. We also assign the probability $\frac{1-\alpha}{2N_2}$ to choose a particular null flip, and the probability $\frac{1-\alpha}{2N_3}$ to choose a particular positive flip. As a consequence the probability to choose any null flip is $\frac{1-\alpha}{2}$, and the probability to choose any positive flip is also $\frac{1-\alpha}{2}$. If there are no positive flips available, $N_3 = 0$ , then we simply take the probability to choose any null flip $1 - \alpha$ (distributed equally among the null flips available) . If there are no null flips available, $N_2 = 0$ , then we simply take the probability to choose any positive flip $1 - \alpha$ (distributed equally among the positive flips available).
If there are no negative flips available , $N_1 = 0$ , then we choose a null or positive flip with equal probability $\frac{1}{2}$ (distributed equally among the flips available). It is, in principle, possible that at a given moment we only have flips of a certain type available. In that case, the algorithm will choose that type of flip with probability 1.

So let's see what is happening. At each step, the algorithm will choose null or positive flips, with high probability (we assume α close to 0), so the number of satisfied clauses can only increase, with high probability (and any negative flips that may occur will be reversed later with high probability). We have to prove that the positive flips occur sufficiently often, so the algorithm is polynomial. This point will be discussed in a future paper.

**The clustered Sparrow algorithm**

*We start with a 3SAT expression E. We construct the clustered expression E*, where each variable appears in at most 3 clauses, as described above.*

*Repeat* $2m^2$ *times (m is the number of clauses of E*)*

> *If the current truth assignment is a solution, return the solution and stop*
>
> *Choose at random a negative, null or positive flip, according to the probability function described above.*
>
> *Perform the flip, and update.*

*Return no solution was found.*

So at every step, the algorithm looks only at all the variables that have unsatisfied clusters, and chooses a negative, null or positive flip, according to the probability function described above.

The make – break idea is discussed in [Schöning, 2012]. The idea to work with clustered expressions E*, as well as using a probability function that takes into account all the variables at every step (not just those involved in an unsatisfied clause) is original. This algorithm is the result of my collaboration with Dr. Anastasia-Maria Leventi-Peetz (BSI, Germany), but implementation and testing have not been performed yet.

Cristian Dumitrescu,
119 Young St., Ap. 11,
Kitchener, Ontario N2H 4Z3,
Canada.

Email: cristiand43@gmail.com
cristiand41@hotmail.com

Tel : (519) 574-7026